%% file: main.tex
\documentclass{article}
\usepackage[english]{babel}

\usepackage[letterpaper,top=2cm,bottom=2cm,left=3cm,right=3cm,marginparwidth=1.75cm]{geometry}

\usepackage{amsmath}
\usepackage{listings}
\usepackage{graphicx}
\usepackage{caption}
\usepackage{subcaption}
\usepackage{color}
\usepackage{tikz}
\usepackage[utf8]{inputenc} 
\usepackage[T1]{fontenc}
\usepackage{wrapfig}
\usepackage[colorlinks=true, allcolors=blue]{hyperref}

\lstdefinestyle{customasm}{
    belowcaptionskip=1\baselineskip,
    frame=single, 
    frameround=tttt,
    xleftmargin=\parindent,
    language=[x86masm]Assembler,
    basicstyle=\footnotesize\ttfamily,
    commentstyle=\itshape\color{green!60!black},
    keywordstyle=\color{blue!80!black},
    identifierstyle=\color{green!80!black},
    tabsize=4,
    numbers=left,
    numbersep=8pt,
    stepnumber=1,
    numberstyle=\tiny\color{gray}, 
    columns = fullflexible,
}

\title{Branch Target Buffer Reverse Engineering on Arm}
\author{Junpeng Wan}

\begin{document}
\maketitle




\section{Introduction}
The Branch Target Buffer (BTB) plays a critical role in efficient CPU branch prediction. Understanding the design and implementation of the BTB provides valuable insights for both compiler design and the mitigation of hardware attacks such as Spectre. However, the proprietary nature of dominant CPUs, such as those from Intel, AMD, Apple, and Qualcomm, means that specific BTB implementation details are not publicly available. To address this limitation, several previous works (e.g., \cite{btbrex86, yavarzadeh2023half}) have successfully reverse-engineered BTB information, including capacity and associativity, primarily targeting Intel's x86 processors. However, to our best knowledge, no research has attempted to reverse-engineer and expose the BTB implementation of ARM processors.

This project aims to fill the gap by exploring the BTB of ARM processors. Specifically, we investigate whether existing reverse-engineering techniques developed for Intel BTB, as outlined in \cite{btbrex86}, can be adapted for ARM. We reproduce the x86 methodology and identify specific PMU events for ARM to facilitate the reverse engineering process. In our experiment, we investigated our ARM CPU, i.e., the quad-core Cortex-A72 of the Raspberry Pi 4B. Our results show that the BTB capacity is 4K, the set index starts from the 5th bit and ends with the 15th bit of the PC (11 bits in total), and there are 2 ways in each set. The source code can be find at \href{https://github.com/stefan1wan/BTB_ARM_RE}{https://github.com/stefan1wan/BTB\_ARM\_RE}.

\input{docs/Background.tex}
\input{docs/Methodology.tex}

\input{docs/EXP.tex}

\section{Summary and Future Work}
In this work, we reverse-engineered the BTB of the Raspberry Pi 4B, determining the capacity to be 4K, the set index to span from the 5th bit to the 15th bit of the PC, and the associativity to be 2. 
Building on this, the exploration of Apple's M-series processors represents a promising direction for future research.

\bibliographystyle{unsrt}
\bibliography{main}

\end{document}

%% file: docs/Background.tex
\section{Background and Related Work}

\subsection{Branch Target Buffer(BTB)}

The Branch Target Buffer (BTB) is a critical component of the Branch Prediction Unit (BPU) that optimizes the processing of branching operations. It stores the history of branch target program counter (PC) addresses, enabling the processor to quickly access and execute the target instruction when needed.
Consider the following example: when the processor executes the instruction $0x804000: jmp \\ ptr [0x80a000]$, it typically fetches the address stored at $0x80a000$ and then executes the subsequent instructions. However, with a BTB in place, the BPU can search for the branch target of $0x804000$ and jump directly to it, thereby reducing the time spent parsing the target.

It is worth noting that the BTB supports conditional branches in addition to unconditional branches, as illustrated in this example. Furthermore, it is important to distinguish between Branch Target Prediction and Branch Prediction. The former predicts the precise branch target PC address and relies on additional information, such as addresses stored in the BTB, while the latter predicts whether the branch will be taken or not. Similar to other caches (such as L1, L2, and LLC), the BTB is organized using sets and ways, with indexes used to refer to sets and tags used to identify specific PC addresses.
    
\subsection{BTB Reverse Engineering on x86}
There is no public information available regarding the BTB capacity for most CPUs, including the one under investigation. Previous work has proposed a solution for reverse-engineering the BTB on the Intel Pentium M \cite{btbrex86}. In this project, we aim to adapt and reproduce their methodology for the ARM architecture.

%% file: docs/Methodology.tex
\section{Experiment}

\subsection{Platform and PMUs}
The target platform for this project is a Raspberry Pi 4 (Model B), which features Quad core Cortex-A72 (ARM v8) CPU~\cite{raspberrypi4b}. The operating system is Ubuntu 22.04.5 LTS with linux kernel version 5.15.0-1064-raspi. 

Performance Monitoring Units (PMUs) in ARMv8-A processors are essential for performance profiling, providing insights into metrics like CPU cycles and instruction counts. By default, access to PMU counters is restricted to privileged execution levels (EL1 or higher) to maintain system security and stability. However, enabling user-level (EL0) access can facilitate efficient performance analysis within user-space applications. To achieve this, we instert  a kernel module  that modifies control registers, such as \texttt{pmuserenr\_el0} and \texttt{pmcntenset\_el0}, to permit EL0 access to PMU counters~\cite{el0_pmu}. This approach allows user-space applications to utilize PMU data without extensive kernel modifications, thereby supporting effective profiling across ARMv8-based platforms.
In this project, we specifically use a PMU event called \textit{ARM\_PMU\_BR\_MIS\_PRED}, which counts the number of branch mispredictions~\cite{arm2015armv8}.





\subsection{BTB Organization}  
\begin{figure}
    \centering
    \includegraphics[width=0.5\textwidth]{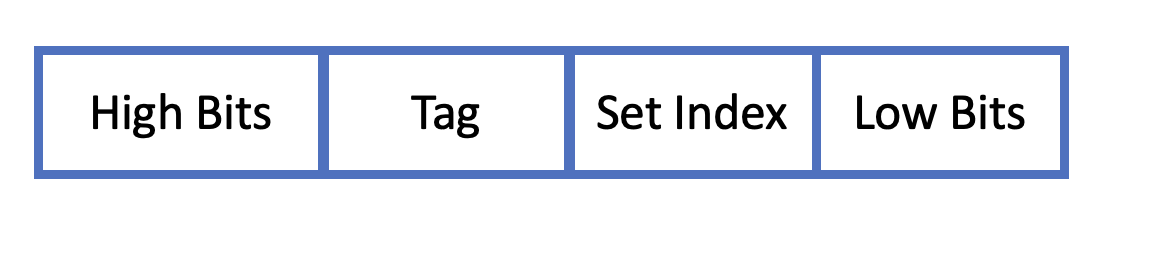}
    \caption{\label{fig:btbaddr}BTB-related virtual address bits.}
\end{figure}

We need to make some assumptions about the BTB organization before starting the reverse engineering work. Similar to other types of caches, 
the BTB has a capacity of $C = S \times W$, where $S$ is the number of sets, and $W$ is the number of ways in each set~\cite{btbrex86}. 
In this work, we also assume that some lower bits of the PC address are used for the set index, while some higher bits are used for tags, as shown in Figure~\ref{fig:btbaddr}. Besides, there might be some extra lowest bits and highest bits that are not used for the set index or the tag, as also shown in Figure~\ref{fig:btbaddr}.

\subsection{Methodology}

The basic concept behind the reverse engineering method is straightforward: we execute $B$ branches over many rounds and count the number of branch mispredictions. When the number of branch mispredictions is low, it indicates that the BTB buffers all the branch targets. Conversely, when the number of branch mispredictions is high or approaches 100\% rate, it suggests that the BTB has either reached its capacity, certain sets are not being utilized, or there is a tag conflict within a cache set. Besides $B$, we also vary the distance between two adjacent branches by filling the gap with \texttt{nop} operations. The gap, denoted as $N$, is aligned to powers of 2, such as 8, 16, or 32. This alignment fixes certain bits to 0, allowing us to explore behavior related to the tag and set index. 

Following the above description, we develop a basic gadget that contains $B$ indirect branches, with the address gap between them denoted as $N$. We make both $B$ and $N$ flexible to enable exploration of various scenarios. In this gadget, unconditional indirect branches are adopted to ensure the utilization of the Branch Target Buffer (BTB) while avoiding branch mispredictions unrelated to the BTB. To use indirect branches, we employ two ARM instructions: the \texttt{adr} instruction to load the label of the next branch and the \texttt{br} instruction to jump to that label. Since these two instructions together occupy 8 bytes (in ARMv8, each instruction is 4 bytes), we start analyzing $N$ from 8, i.e., with 0 \texttt{nop} instructions in between. By counting branch mispredictions using PMU events, we can roughly determine branch target misses within the BTB during the execution of this gadget. The pseudocode of the gadget is shown in Listing~\ref{ARMcode}.

\begin{lstlisting}[style=customasm, caption={Test Gadget Pseudocode}, label=ARMcode]
\label{listing:gadget}
0: adr x0, next1
4: BR X0
8: nop
12: nop
.......
 
next1:
N: adr x0, next2
N+4: BR X0
N+8: nop
N+12: nop
.......
 
next2:
2N: adr x0, next3
2N + 4: BR X0
2N + 8: nop
2N + 12: nop
.......

(anothor B-3 blocks)

nextB:
B*N:ret

\end{lstlisting}


In the following steps, we measure branch misprediction rates for different values of $B$ and $N$. For each test with the same $B$ and $N$, we follow these steps: (1) warm up the BTB by executing the test gadget 10 times, and (2) run the test gadget once and count the number of branch mispredictions ($C$). The BTB miss rate is then calculated as $C / B$. By comparing and analyzing the misprediction rates for all values of $B$ and $N$, we explore the characteristics of the BTB and BPU, as discussed in Subsection~\ref{subsec:ana}.

%% file: docs/EXP.tex
\subsection{Analysis and Result}
\label{subsec:ana}

\begin{figure}[h]
     \centering
     \includegraphics[width=0.6\textwidth]{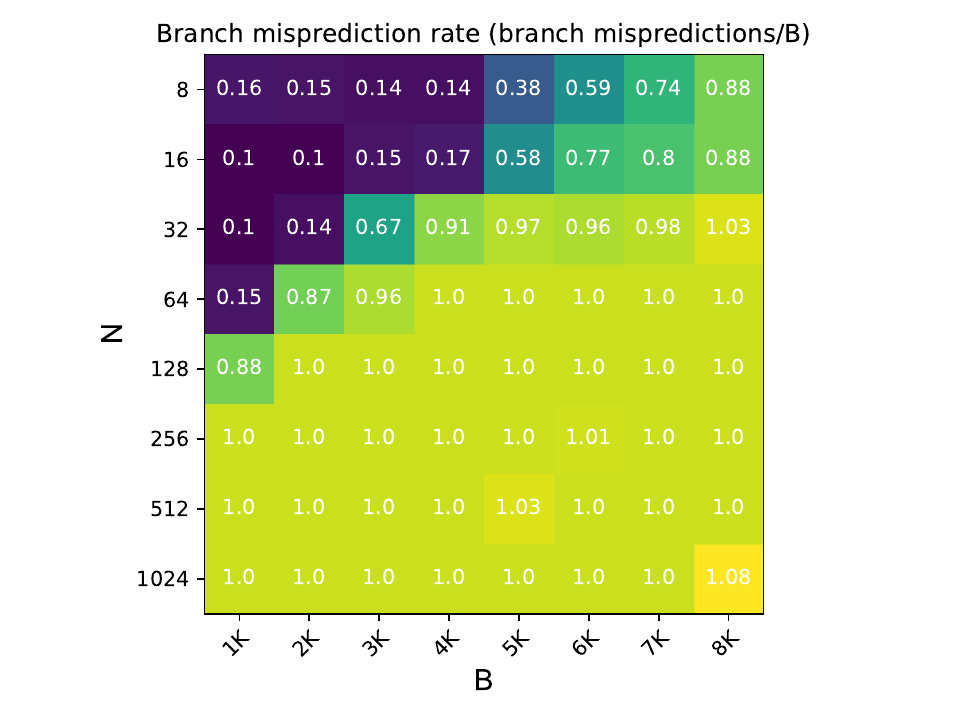}
      \caption{BTB Capacity Reverse Engineering}
      \label{fig:capacity_arm}
\end{figure}

\noindent\textbf{Capacity.} To reverse engineer the capacity, we vary $B$ from 1K to 8K and $N$ from 8 to 1024 for analysis. The results obtained are shown in Figure~\ref{fig:capacity_arm}. In this figure, we present a heatmap to visualize the BTB miss rates, where cooler colors represent lower miss rates and warmer colors represent higher miss rates. Note that in some blocks, the branch misprediction rate is slightly larger than 1, possibly due to branch mispredictions caused by interrupts. We do not address this noise issue currently, as it does not affect our analysis. We can infer the following information based on the information in the figure. 

\begin{itemize}
\item  The BTB set index starts from the 5th bit, and the reasoning is as follows: When N is 8 or 16  and B increases from 1K to 8K, the misprediction rate remains similar.  However, when N becomes 32, the misprediction rate suddenly increases, especially when B is between 3K and 8K. This increase in mispredictions occurs because, if N is 32 (or if the first 5 bits are fixed to 0), the number of BTB sets being used is effectively halved.  This indicates that the 5th bit of PC is included in the set index. Besides, the 4rd is not included, or else the misprediction rate will be different for $N = 8$ and $N = 16$.  There for we infer that the BTB set index starts from the 5th bit.

\item The BTB entry size is 4K (4096). The reasoning is as follows: since the set index starts from the 4th bit, we focus on the second line of observations, i.e., when N is 16. The misprediction rate is 0.17 for 4K entries, whereas it increases significantly to 0.48 for 5K entries. We interpret the 0.17 as a low misprediction rate, close to zero, with any additional variations likely caused by noise factors, such as timing interruptions. 
\end{itemize}

In this way, our reverse engineering results indicate that the BTB capacity for the Raspberry Pi 4B is 4K, and the set index starts from the 5th bit. Moreover, the effectiveness of our results on the BTB capacity aligns with the actual data ~\cite{CortexA72FetchBranch}.

\begin{figure}[h]
     \centering
     \includegraphics[width=0.6\textwidth]{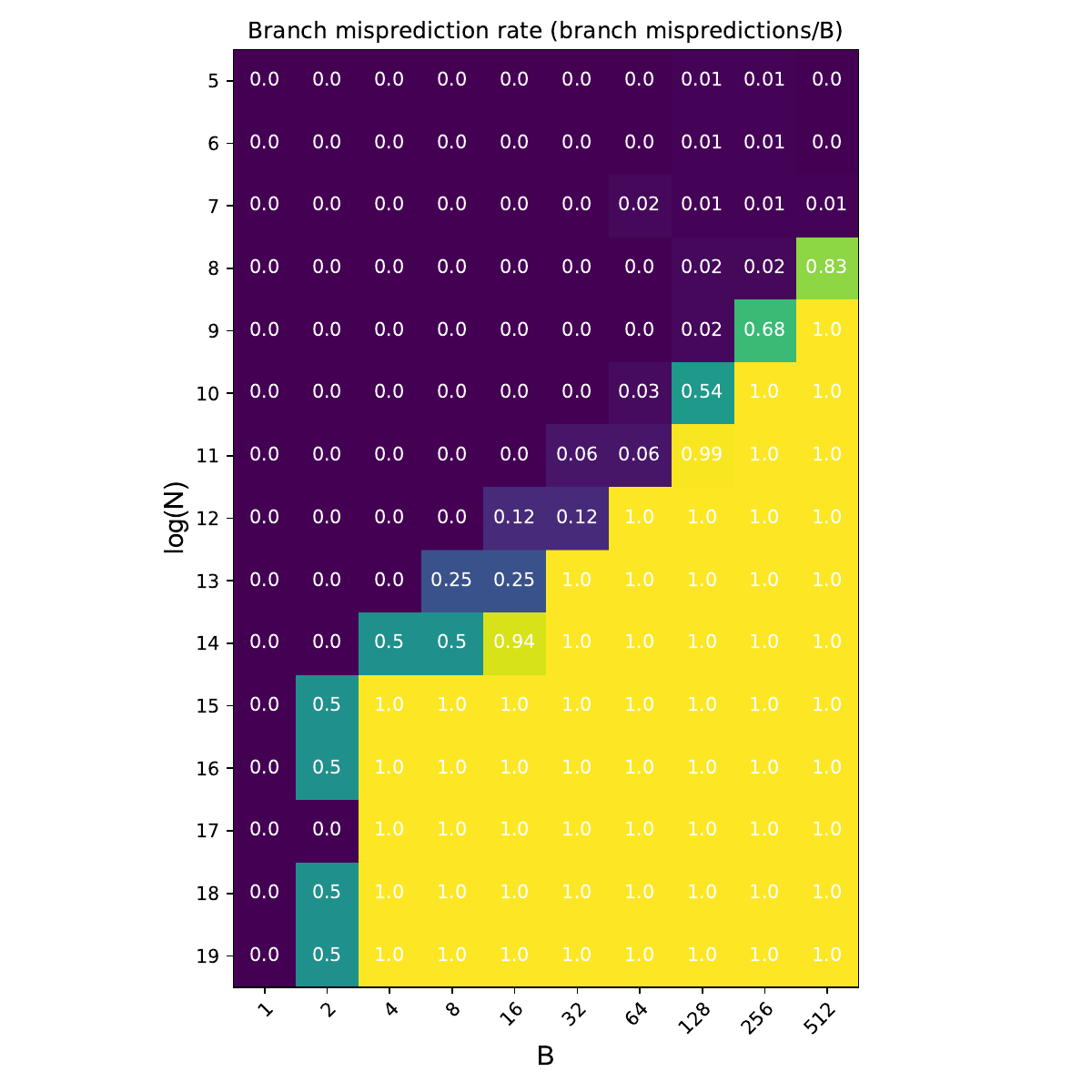}
      \caption{BTB Set Index and Ways Reverse Engineering}
      \label{fig:setindex_arm}
\end{figure}

\noindent\textbf{Set Index and Ways.}  To reverse engineer the set index, we vary $B$ from 1 to 512 and $N$ from $2^{5}$ to $2^{19}$ for analysis. The results obtained are shown in Figure~\ref{fig:setindex_arm}. From the figure, we can see that when $N \geq 2^{15}$, the misprediction rate remains similar for each $B$. Moreover, $N = 2^{15}$ implies that the first 15 bits of the branch address are set to 0. Thus, we infer that for $N \geq 2^{15}$, the fixed bits exceed all the set index bits, and only one BTB cache set is utilized. In this way, we conclude that the BTB set index ends at the 15th bit. Combined with previous results, the set index starts from the 5th bit and ends at the 15th bit, resulting in 11 bits to index the cache sets. This indicates that there are at most 2048 ($2^{11}$) cache sets in the BTB. Consequently, there must be at least 2 ways in each cache set.

Moreover, when $N \geq 2^{15}$ and $B = 2$, the miss rate is 0 or 1, indicating that 0 or 1 branches are mispredicted. Notably, when $N = 2^{17}$, the miss rate is 0 for $B = 2$, leading us to infer that there are at least two ways. Additionally, when $N \geq 2^{15}$ and $B \geq 4$, the miss rate stabilizes at 1, suggesting that there are exactly two ways. In summary, the BTB has 2048 sets, indexed by 11 bits (5th to 15th bit), and 2 ways in each set.

